\documentclass{aastex63}

\shorttitle{Local Group distances and publication bias. VI.}
\shortauthors{Richard de Grijs and Giuseppe Bono}

\begin{document}

\title{Clustering of Local Group distances: publication bias or
  correlated measurements? VI. Extending to Virgo cluster distances}

\author{Richard de Grijs}
\affiliation{Department of Physics \& Astronomy, Macquarie
  University, Balaclava Road, Sydney, NSW 2109, Australia}
\affiliation{Research Centre for Astronomy, Astrophysics \&
  Astrophotonics, Macquarie University, Balaclava Road, Sydney, NSW
  2109, Australia}
\affiliation{International Space Science Institute--Beijing, 1
  Nanertiao, Zhongguancun, Hai Dian District, Beijing 100190, China}

\author{Giuseppe Bono}
\affiliation{Dipartimento di Fisica, Universit\`a di Roma Tor
  Vergata, via Della Ricerca Scientifica 1, 00133, Roma, Italy}
\affiliation{INAF, Rome Astronomical Observatory, via Frascati
  33, 00078 Monte Porzio Catone, Italy}

\begin{abstract}
We have established an internally consistent Local Group distance
framework, using the Galactic Center, the Large Magellanic Cloud, and
Messier 31 (M31) as important stepping stones. At greater distances,
few distance benchmarks are available. As a consequence, M87 and/or
Virgo cluster distances are often invoked as the next rung on the
ladder to more distant objects such as the Fornax and Coma
clusters. Therefore, we extensively mined the published literature for
independently derived distance estimates to either M87 or the center
of the Virgo cluster. Based on our newly compiled, comprehensive
database of 213 such distances, published between 1929 and 2017 July,
we recommend an outward extension to our distance framework,
$(m-M)_0^{\rm M87} = 31.03 \pm 0.14$ mag ($D = 16.07 \pm 1.03$ Mpc;
where the uncertainty represents the Gaussian $\sigma$ of the
distribution), based on a subset of recent (post-1990) M87/Virgo
cluster distance measurements. The most stable distance tracers
employed here were derived from analysis of both primary and secondary
distance indicators. Among the former, we preferentially rely on
Cepheid period--luminosity relations and red-giant-branch terminal
magnitudes; our preferred secondary distance tracers are surface
brightness fluctuations. Our updated distance modulus to M87 implies a
slightly reduced black hole mass of $(5.9 \pm 0.6) \times 10^9
M_\odot$ with respect to that determined by the Event Horizon
Telescope collaboration.
\end{abstract}


\section{Beyond the Local Group: Messier 87 and the Center of the Virgo Cluster}

Triggered by the suggestion that published distance measurements to
the Large Magellanic Cloud (LMC) may have been subject to confirmation
or publication bias (Schaefer 2008), earlier this decade we set out to
explore that concern in great depth. Although we were unable to
confirm Schaefer's (2008) suggestion on the basis of the most
comprehensive database of LMC distance moduli collected at the time
(de Grijs et al. 2014, henceforth Paper I), we managed to put Freedman
et al.'s (2001) canonical LMC distance modulus of $(m-M)_0^{\rm LMC} =
18.50 \pm 0.10$ mag on a much more robust statistical footing, i.e.,
yielding $(m-M)_0^{\rm LMC} = 18.49 \pm 0.09$ mag (independently
confirmed by Crandall \& Ratra 2015, using the same data set).

Next, we decided to take our analysis of possible publication bias in
distance estimates to Local Group galaxies to the next level. We
expanded our scope to include Messier 31 (M31) and a number of its
larger and better-studied companion galaxies, i.e., M32, M33, NGC 147,
NGC 185, NGC 205, IC 10, and IC 1613 (de Grijs \& Bono 2014: Paper
II), and the Small Magellanic Cloud (SMC; de Grijs \& Bono 2015: Paper
III; see for independent confirmation based on the same data set,
Crandall \& Ratra 2015). The latter revealed significant difficulties
related to the definition of `the' SMC center, given the galaxy's
irregular morphology, as well as important depth effects reflected
differently by different distance tracers.

We proceeded to tie this updated Local Group distance scale to one of
the most important and nearest galactic-scale distances, the distance
to the Galactic Center (de Grijs \& Bono 2016: Paper IV), for which we
recommended $R_0 = 8.3 \pm 0.2 \mbox{ (statistical)} \pm 0.4$
(systematic) kpc (equivalent to $(m-M)_0^{\rm Gal.C.} = 14.60 \pm
0.05$ mag, where the uncertainties reflect the statistical
uncertainties only), with an associated Galactic rotation constant at
the solar Galactocentric radius, $\Theta_0 = 225 \pm 3 \mbox{
  (statistical)} \pm 10$ (systematic) km s$^{-1}$ (de Grijs \& Bono
2017: Paper V). In very recent developments, the Gravity Collaboration
et al. (2019) published an updated geometric distance determination to
the Galactic Center based on 27 years of orbital measurements for the
star S2, yielding $R_0 = 8178 \pm 13 \mbox{ (statistical)} \pm 22$
(systematic) pc. Our multiple-tracer statistical determination from
2016 remains fully commensurate with this latest geometric distance
determination.

Thus, through careful meta-analysis of published distance measures, we
established an internally consistent and statistically supported Local
Group distance framework, using the Galactic Center, the LMC, and
M31---$(m-M)_0^{\rm M31} = 24.46 \pm 0.10$ mag (Paper II)---as
important stepping stones. In addition, we advocated the use of
NGC~4258 as the next step in the distance hierarchy on account of its
water maser-based geometric distance determination (see the discussion
in Paper II).

Not surprisingly, published distance measurements to galaxies beyond
the Local Group become sparser with increasing distance. A notable
exception in this context are distances to the Virgo cluster and its
dominant giant elliptical galaxy, Messier 87 (M87). At this distance,
we have access to few (if any) alternative distance benchmarks, and as
a consequence M87/Virgo cluster distances are often invoked as
stepping stone to more distant objects such as the Fornax and Coma
clusters. (We will return to this latter aspect in Section
\ref{discussion.sec}.) Therefore, in this paper we set out to obtain a
firm distance estimate to M87 or, alternatively, to the center of the
Virgo cluster, based on a similar meta-analysis of published distance
measures (and Galactic rotation constants) as we undertook for Papers
I through V.

Before we set out to mine the literature, however, we needed to
consider the three-dimensional (3D) spatial distribution of the
galaxies in the Virgo cluster. It is well established that the
cluster's mass distribution is not smooth but distributed across a
number of subclumps. Most notably, NGC 4697 is the central elliptical
galaxy in the foreground `NGC 4697 group,' which also includes NGC
4731 and a number of smaller galaxies in the Virgo Southern
Extension. On the other hand, the W$'$ group is located behind the
main body of the Virgo cluster (e.g., Cantiello et al. 2018; their
Table 2). The majority of Virgo cluster member galaxies are
distributed in the so-called A, B, and E subclusters, where subclump B
is thought to be located 0.4--0.5 mag (in distance modulus) behind
subclumps A (e.g., Yasuda et al. 1997; Federspiel et al. 1998;
Feldmeier et al. 1998)---which is associated with M87 (e.g., Gavazzi
et al. 1999)---and E (Gavazzi et al. 1999; Neilsen \& Tsvetanov 2000).

Current consensus suggests that M87 is located near the physical
center of the Virgo cluster (e.g., Ciardullo et al. 1998; Mei et
al. 2007; Blakeslee et al. 2009; Bird et al. 2010), despite the
well-known radial velocity discrepancy of $\sim 200$ km s$^{-1}$
between M87 and the Virgo cluster mean (Binggeli et al. 1987), with
M87 exhibiting the higher velocity, and its projected $\sim 1^\circ$
distance from the center of the Virgo cluster's isopleths. Ciardullo
et al. (1998), following Bird (1994), argued that such velocity and
positional offsets are common in central elliptical galaxies in
dynamically young galaxy clusters---including the Virgo cluster.

Nevertheless, it is not straightforward to define the Virgo cluster's
center even when using only galaxies associated with the main
subclusters tracing the Virgo cluster's potential well. Elliptical and
early-type galaxies are more spherically distributed than the
cluster's complement of spiral galaxies. The latter are distributed in
an elongated structure extending from 13 to 30 Mpc (Planck
Collaboration et al. 2016) along the line of sight. This affects the
determination of average cluster distance moduli based on diagnostics
typical of either late- (Tully--Fisher relation) or early-type
(Faber--Jackson/$D_n-\sigma$ analysis) differently (e.g., Fukugita et
al. 1993).

In Section \ref{data.sec}, we outline our approach to obtaining our
data set of published distance moduli to our target region. Then, in
Section \ref{analysis.sec} we consider the statistics of the resulting
distance measurements for individual tracer populations, which we
discuss in Section \ref{discussion.sec}. We also summarize and
conclude the paper in Section \ref{discussion.sec}.

\section{Data mining}
\label{data.sec}

To compile a comprehensive database of distances to either M87 or the
center of the Virgo cluster, we extensively perused the
NASA/Astrophysics Data System (ADS). As of 2019 September 3, a search
for `M87' returned 7922 hits. Our manual perusal resulted in an
overall tally of 213 independently derived distance values, spanning
the period from Hubble's (1929) determination based on a comparison of
galaxy luminosities to Hartke et al.'s (2017) reanalysis of the
planetary nebulae population associated with Messier 49, published in
July 2017.

The NASA/ADS database contains all articles published in the main
astrophysics journals since 1975. Its coverage of historical records
is continuously increasing. In this paper, we will focus predominantly
on the most modern M87/Virgo cluster distance determinations. In
practice, this means that our analysis will be based on post-1985
measurements. We are confident that we have tracked down the vast
majority of such measurements published in the literature.

\begin{figure*}[ht!]
\vspace{-3cm}
\plotone{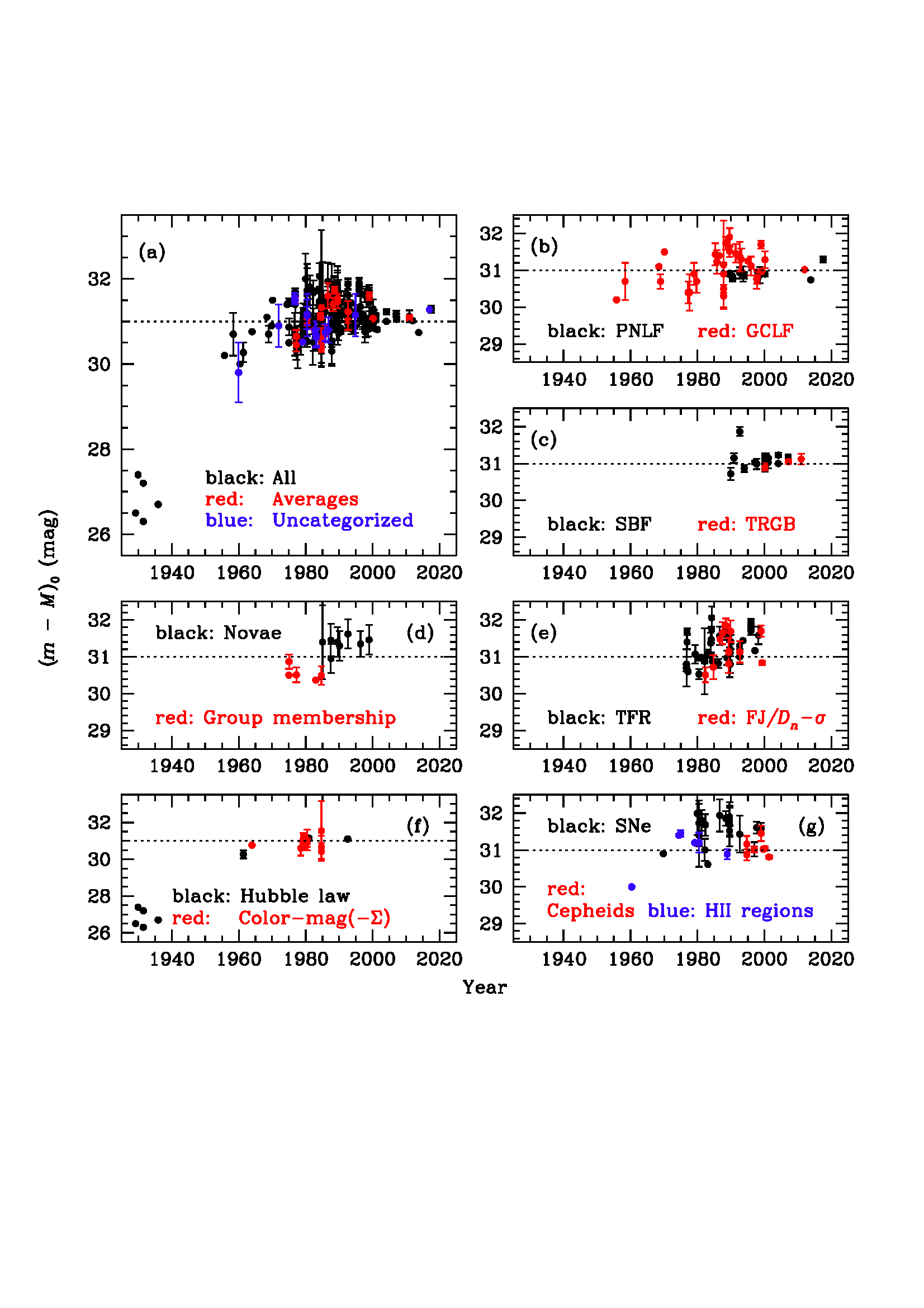}
\vspace*{-4.5cm}
\caption{Published distances to either M87 or the center of the Virgo
  cluster as a function of publication date (original values with
  their original error bars, where available). The horizontal dotted
  lines are drawn at $(m-M)_0 = 31.0$ mag and are meant to guide the
  eye. $D_n-\sigma$: Edge-on projection of the Fundamental Plane of
  elliptical galaxies, where $D_n$ is the diameter within which the
  effective surface brightness is 20.75 $\mu_B$ ($B$-band surface
  brightness) and $\sigma$ is the mean velocity dispersion within the
  galaxy's effective radius; FJ: Faber--Jackson relation; GCLF:
  Globular cluster luminosity function; PNLF: Planetary nebula
  luminosity function; SBF: Surface brightness fluctuations; SNe:
  Supernovae; TFR: Tully--Fisher relation; TRGB: Tip of the red giant
  branch. `Averages' in panel (a) include weighted and unweighted
  means of different methods of distance determination to M87, as well
  as mean values of the distance moduli to samples of central Virgo
  Cluster galaxies, as published by the original authors (see Section
  9 in our externally linked data table); `Uncategorized' distance
  moduli include any measurements that are not already included in the
  other panels, mostly because of the scarcity of data points for a
  particular measurement approach.}
\label{fig1}
\end{figure*}

We have kept careful track of the provenance of our distance
estimates, whether based on M87 measurements or pertaining to the
Virgo cluster more broadly. Where possible, we have only retained
Virgo cluster measures that relate to the M87 subcluster, or which
were centered on M87, for further analysis. Our final database, sorted
by year and by tracer, is available online through
http://astro-expat.info/Data/pubbias.html,\footnote{A permanent link
  to this page can be found at
  http://web.archive.org/web/20160610121625/http://astro-expat.info/Data/pubbias.html;
  members of the community are encouraged to send us updates or
  missing information.} Figure \ref{fig1}a shows the overall data set
in black; published weighted average values are overplotted in
red. Panels (b) through (g) show our sample of distance measurements
split by tracer. Where available, we have included the published
$1\sigma$ error bars (random errors only in case systematic errors
were also published) as well as horizontal dotted lines to guide the
eye, located at $(m-M)_0 = 31.0$ mag. At first glance, none of the
tracers, nor the data set as a whole, exhibit any systematic
trends. However, it is clear that some tracers are subject to
significantly greater scatter than others, and systematic differences
in mean levels are seen when comparing different tracers. We will
discuss these issues in the next section.

\section{Differences among tracers}
\label{analysis.sec}

Figure \ref{fig1} shows that for a number of tracers we only have
access to rather old (pre-1990) data. Tracers in this category include
distance determinations based on group membership, Hubble's
velocity--distance law, color--magnitude analysis and related methods,
H{\sc ii} region sizes, and also supernova (SN)-based
distances. Others show spreads that significantly exceed the published
statistical error bars (thus suggesting the presence of
unaccounted-for systematics), including distance determinations based
on the globular cluster luminosity function (GCLF) and dynamical
distance estimates (e.g., those based on the Tully--Fisher or
Faber--Jackson relations and the $D_n-\sigma$ projection of the
Fundamental Plane of elliptical galaxies). Only five data sets appear
internally consistent, leading to fairly tight averages: those based
on Cepheids, the planetary nebulae luminosity function (PNLF), surface
brightness variations (SBF), the tip of the red-giant branch (TRGB)
magnitude, and novae (although the error bars associated with the
latter are large).

\begin{figure}[ht!]
\epsscale{1.2}
\vspace{-1.5cm}
\plotone{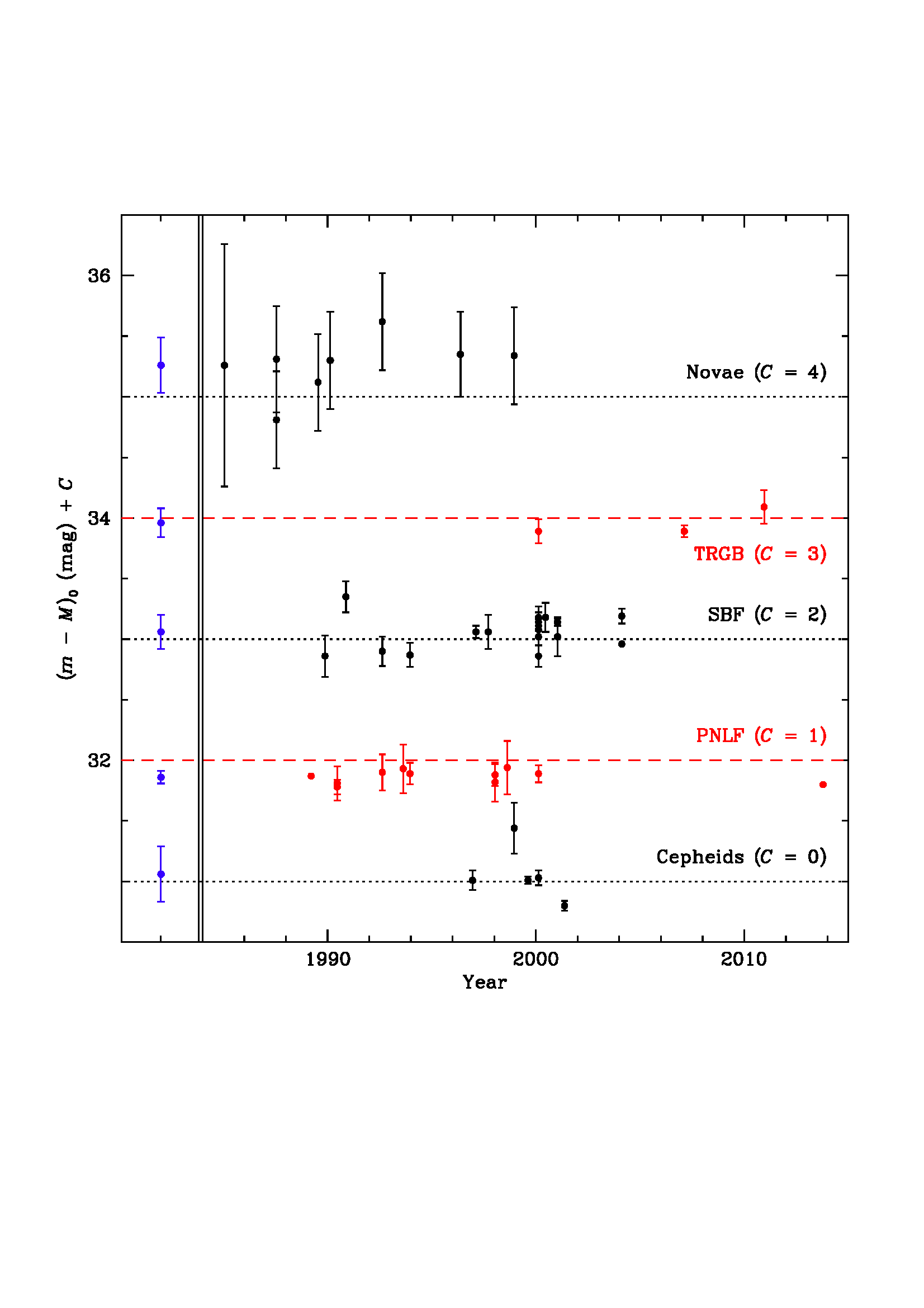}
\vspace{-3cm}
\caption{`Cleaned' M87/Virgo cluster distances and their published
  $1\sigma$ error bars for five data sets that include the most recent
  determinations available, offset by constants steps $C$ for
  clarity. Individual distances have been corrected, where necessary,
  to a common distance scale, i.e., tied to $(m-M)_0^{\rm LMC} =
  18.49$ mag. The blue data points in the left-hand panel represent
  the mean values and their $1\sigma$ uncertainties (assuming Gaussian
  distributions, for simplicity) of the individual tracers. The
  horizontal dotted and dashed lines are drawn at $(m-M)_0 = 31.0 + C$
  mag, and are meant to guide the eye.}
\label{fig2}
\end{figure}

We carefully examined the procedures followed by the original authors
to arrive at each individual distance estimate for these latter
tracers. In addition, we adjusted the published distances to conform
with the distance framework established in the previous papers in this
series, in essence tied to $(m-M)_0^{\rm LMC} = 18.49$ mag. In
practice, these adjustments were primary calibration offsets of order
a few hundredths of a magnitude only, in most cases; see Table
\ref{tab1} for the full adjusted data set. The resulting homogenized
distance measurements are shown in Figure \ref{fig2} as a function of
distance indicator, where we have added a constant $C \in [0,4]$ mag
to offset the individual tracers from one another for reasons of
clarity. Note that the distance moduli for M87 and the Virgo Cluster
(see Table \ref{tab1}, third column, for this distinction) are
statistically indistinguishable, so that henceforth we will base our
analysis on the combined data set.

\begin{table*}
\caption{Adopted `adjusted' distance moduli used in this paper.}
\label{tab1}
\begin{center}
{\scriptsize
\tabcolsep 0.5mm
\begin{tabular}{@{}ccccll@{}}
\hline \hline
Publ. date & $(m-M)_0$ && M87 (`M') or & \multicolumn{1}{c}{Reference} & \multicolumn{1}{c}{Notes} \\
(mm/yyyy) & (mag) && Virgo (`V') \\
\hline
\multicolumn{6}{c}{1. Cepheids}\\
\hline
12/1996 & $31.01 \pm 0.08$ && V & van den Bergh (1996) \\
12/1998 & $31.44 \pm 0.21$ && V & Tammann et al. (2000) \\
08/1999 & $31.01 \pm 0.03$ && V & Macri et al. (1999) \\
02/2000 & $31.03 \pm 0.06$ && M & Ferrarese et al. (2000) \\
05/2001 & $30.80 \pm 0.04$ && V & Freedman et al. (2001) & Recalibration \\
\hline
\multicolumn{6}{c}{2. PNLF}\\
\hline
03/1989 & $30.87 \pm 0.  $ && V & Jacoby (1989) \\
06/1990 & $30.81 \pm 0.14$ && V & Jacoby et al. (1990) \\
06/1990 & $30.78 \pm 0.06$ && M & Jacoby et al. (1990) \\
08/1992 & $30.90 \pm 0.15$ && V & Jacoby et al. (1992) \\
08/1993 & $30.93 \pm 0.2 $ && V & M\'endez et al. (1993) \\
12/1993 & $30.89 \pm 0.09$ && V & Ciardullo et al. (1993) \\
01/1998 & $30.82 \pm 0.16$ && M & Ciardullo et al. (1998) \\
01/1998 & $30.88 \pm 0.09$ && M & Ciardullo et al. (1998) & Rescaled Jacoby et al. (1990) value \\
08/1998 & $30.94 \pm 0.22$ && M & Feldmeier et al. (1998) \\
02/2000 & $30.89 \pm 0.07$ && M & Ferrarese et al. (2000) \\
10/2013 & $30.80 \pm 0.  $ && M & Longobardi et al. (2013) \\
\hline
\multicolumn{6}{c}{3. SBF}\\
\hline
11/1989 & $30.86 \pm 0.17$ && V & Tonry et al. (1989) \\
11/1990 & $31.35 \pm 0.13$ && V & Tonry et al. (1990) & $I$ \\
08/1992 & $30.90 \pm 0.12$ && V & Jacoby et al. (1992) \\
12/1993 & $30.87 \pm 0.10$ && V & Ciardullo et al. (1993) \\
02/1997 & $31.06 \pm 0.05$ && M & Tonry et al. (1997) \\
09/1997 & $31.06 \pm 0.14$ && M & Neilsen et al. (1999) \\
02/2000 & $31.18 \pm 0.04$ && M & Ferrarese et al. (2000) & $I$ \\
02/2000 & $31.08 \pm 0.06$ && M & Ferrarese et al. (2000) & F814W \\
02/2000 & $30.86 \pm 0.09$ && M & Ferrarese et al. (2000) & $K_{\rm s}$ \\
02/2000 & $31.17 \pm 0.10$ && M & Ferrarese et al. (2000) & $K'$ \\
02/2000 & $31.14 \pm 0.03$ && V & Tonry et al. (2000) \\
02/2000 & $31.02 \pm 0.16$ && M & Tonry et al. (2000) \\
06/2000 & $31.18 \pm 0.12$ && V & Neilsen \& Tsvetanov (2000) \\
01/2001 & $31.14 \pm 0.03$ && V & Tonry et al. (2001) \\
01/2001 & $31.02 \pm 0.16$ && M & Tonry et al. (2001) \\
02/2004 & $31.19 \pm 0.06$ && V & Jerjen et al. (2004) \\
02/2004 & $30.96 \pm 0.  $ && M & Jerjen et al. (2004) \\
\hline
\multicolumn{6}{c}{4. TRGB}\\
\hline
02 2000 & $30.89 \pm 0.10$ && M & Ferrarese et al. (2000) \\
02/2007 & $30.89 \pm 0.05$ && V & Williams et al. (2007) \\
12/2010 & $31.09 \pm 0.14$ && M & Bird et al. (2010) \\
\hline
\multicolumn{6}{c}{5. Novae}\\
\hline
01/1985 & $31.26 \pm 1.0 $ && M & Pritchet \& van den Bergh (1985) & $30.4 < (m-M)_B < 32.4$ mag \\
07/1987 & $31.31 \pm 0.44$ && V & Pritchet \& van den Bergh (1987) \\
07/1987 & $30.81 \pm 0.4 $ && V & Pritchet \& van den Bergh (1987) & Using the Cohen (1985) Galactic calibration \\
07/1989 & $31.12 \pm  0.4$ && V & van den Bergh (1989) \\
02/1990 & $31.30 \pm 0.40$ && V & Capaccioli et al. (1990) \\
08/1992 & $31.62 \pm 0.40$ && V & Jacoby et al. (1992) \\
05/1996 & $31.35 \pm 0.35$ && V & Livio (1997) \\
12/1998 & $31.34 \pm 0.40$ && V & Tammann et al. (2000) \\
\hline \hline
\end{tabular}
}
\end{center}
\end{table*}

By comparison with the canonical $(m-M)_0 = 31.0 (+ C)$ mag lines
included to guide the eye, it appears that our sample of novae imply
systematically larger distances, while the PNLF distances are
systematically shorter. However, since the latter are based on the
extremely sharp bright cut in the PNLF, planetary nebulae detections
are biased toward foreground objects, so that the resulting distance
estimates are, in essence, lower limits.

The left-hand panel of Figure \ref{fig2} shows the mean values and
their $1\sigma$ uncertainties (assuming Gaussian distributions, for
simplicity) of the individual tracers; see Table \ref{tab2} for the
numerical values. Note that the error bar associated with the mean
Cepheid distance is artificially enhanced by the inclusion of an
apparently outlying data point published in December 1998. This latter
value, obtained from Tammann et al. (2000),\footnote{Note that this
  value was initially published in December 1998; although the volume
  of conference proceedings was eventulally published in 2000, we
  assigned this data point to its initial publication date since we
  are interested in tracing the evolution of Virgo Cluster/M87
  distance moduli with time.} is part of a data set of galaxy
distances that are commonly referred to as the `long distance scale.'

A `long' versus `short' distance scale debate raged in the second half
of the 20th century, associated with, respectively, low and high
values of the Hubble constant (for a review, see, e.g., de Grijs
2011). However, since the publication of Freedman et al.'s (2001)
seminal study on the extragalactic distance scale, consensus has been
reached that the short distance scale is closer to reality. If we were
to remove the Tammann et al. (2000) data point, the Cepheid average
value for M87 would be reduced to $(m-M)_0 = 30.96 (\sigma = 0.11)$
mag.

By combining the 28 data points for the Cepheids, TRGB, and SBF
measurements, which are all well-calibrated and independently
determined, the resulting true distance modulus for the Virgo cluster
center is $(m-M)_0 = 31.05$ mag, with a Gaussian spread of $\sigma =
0.16$ mag ($D = 16.2 \pm 1.2$ Mpc). The equivalent values for the data
set without the Tammann et al. (2000) measurement are $(m-M)_0 =
31.03$ mag and $\sigma = 0.14$ mag. None of these tracers suggest any
trend as a function of publication date, and hence taking a straight
average seems justified, also in view of the relatively small number
of data points available.

Our recommended final value is, in fact, fully consistent with almost
all of the weighted average distance moduli to either M87 or the
center of the Virgo cluster published since 1990, with the notable
exception of the Tammann et al. (2000) value, $(m-M)_0 = 31.60 \pm
0.09$ mag. Jacoby et al. (1992) considered seven of the most reliable
extragalactic distance tracers---the GCLF, novae, SNe Type Ia, the
TFR, the PNLF, SBF, and the $D_n-\sigma$ relation---to derive a
weighted average of $(m-M)_0 = 31.02 \pm 0.22$ mag for the Virgo
cluster. 

The correspondence between our final distance modulus and that
advocated by Jacoby et al. (1992) is perhaps somewhat surprising,
given that the calibrations used for the tracers in common are
somewhat divergent; in practice, however, this divergence results in
cancelling out most of the differences. For instance, Jacoby et
al. (1992) based their Cepheid distance calibration on an adopted LMC
distance modulus of $(m-M)_0 = 18.57$ mag, 0.08 mag larger than our
preferred value; on the other hand, their SBF calibration assumes a
distance to M31 of 770 kpc, some 10 kpc closer (correponding to a
difference of 0.03 mag in distance modulus) than the distance of 780
kpc we recommend to render the Local Group distances internally
consistent.

Ferrarese et al. (2000) found $(m-M)_0 = 31.07 \pm 0.03$ mag using a
wide variety of distance indicators, including Cepheids, the PNLF, the
GCLF, SBF, and the TRGB. Finally, Bird et al. (2010) published a mean
value of the distance modulus to M87 based on the TRGB, the PNLF,
globular cluster sizes and SBF/Cepheid distances, of $(m-M)_0 = 31.06
\pm 0.06$ mag. These latter authors adopted calibration conventions
for the tracers in common with our study that are indeed very close
(within 0.01--0.02 mag in terms of the resulting distance moduli) to
our adopted values.

Considering the overall distribution of distance measures (Figure
\ref{fig1}a), at first glance it appears that the intrinsic spread has
narrowed considerably since the turn of the millennium. This is indeed
reflected in the small $\sigma = 0.14$ mag obtained for our combined
sample, which is dominated by more recent measurements. However, this
apparent tightening of the overall distribution is not a manifestation
of publication bias, nor even of any sustained trends. It is caused by
a shift in focus of the community from one subset of distance
indicators to another. As discussed above, the post-2000 measurements
we have considered to reach our final recommendation included distance
estimates based on Cepheid period--luminosity relations, SBF, and the
TRGB magnitude. Where such measures were available prior to 2000, none
of the tracers suggest a significant broadening of their distribution
toward earlier times.

Before 2000, important subsets of our data points included
measurements that exhibited significant scatter, e.g., GCLF-, novae-,
and SNe-based distance measurements, and distances based on galaxy
dynamics (the Tully--Fisher and $D_n-\sigma$ relations). In addition,
the mean distance modulus implied by our subset of group
membership-based distances is systematically closer, while the
distances resulting from novae, SNe, TFR, and $D_n-\sigma$ analysis
all appear to straddle around means that are systematically more
distant than the canonical $(m-M)_0 = 31.0$ mag level. All of these
latter data points come from our pre-2000 data set, and so these
systematic offsets contribute to the appearance of a broadening of the
overall distribution prior to 2000 compared with more recent
measurements.

Finally, although we have considered the full set of data points
post-2000, including distance estimates to M87 and the center of the
Virgo cluster, there is no evidence of any systematic offsets for any
of our tracer populations at more recent times.

\begin{table}
\vspace{2cm}
\caption{Mean, post-1985 published distance measures to M87 and the
  center of the Virgo cluster as a function of tracer population.}
\begin{center}
\label{tab2}
\begin{tabular}{@{}llcrcccr@{}}
\hline \hline
\multicolumn{1}{l}{Tracer} & \multicolumn{1}{r}{$N$} & \multicolumn{1}{c}{Mean} & \multicolumn{1}{c}{$\sigma$} \\
& & (mag) & (mag) \\
\hline
Cepheids$^a$ &  5 & 31.06 & 0.23 \\
PNLF         & 11 & 30.86 & 0.05 \\
SBF          & 17 & 31.06 & 0.14 \\
TRGB         &  3 & 30.96 & 0.12 \\
Novae        &  8 & 31.26 & 0.23 \\
\hline \hline
\end{tabular}
\end{center}
\flushleft 
$^a$ If we were to ignore the Tammann et al. (2000) data point,
published in December 1998, the mean Cepheid distance modulus reduces
to $(m-M)_0 = 30.96, \sigma = 0.11$ mag.
\end{table}

\section{Implications}
\label{discussion.sec}

Armed with our recommended M87 distance modulus, $(m-M)_0 = 31.03$ mag
($\sigma = 0.14$ mag, equivalent to 1.03 Mpc at the galaxy's distance
of 16.07 Mpc), we are now well placed to extend our previously
established local distance framework to the nearest galaxy
clusters. Our internally consistent distance estimates to the Galactic
Center, the LMC, and M31, combined with the geometric
water-maser-based distance determination to NGC 4258, represent a
tight framework to tie the local distance scale to. Table \ref{tab3}
provides a summary of our preferred reference distances.

\begin{table}
\caption{Recommended local distance framework out to the Virgo cluster.}
\label{tab3}
\begin{center}
\begin{tabular}{@{}lcl@{}}
\hline \hline
\multicolumn{1}{c}{Target}   & $(m-M)_0^{\rm rec.}$ & \multicolumn{1}{c}{Reference} \\
& (mag) \\
\hline
Gal. Center$^a$ & $14.60 \pm 0.05$ & Paper IV \\
LMC       & $18.49 \pm 0.09$ & Paper I   \\
SMC       & $18.96 \pm 0.02$ & Paper III \\
M31       & $24.45 \pm 0.10$ & Paper II  \\
NGC 4258  & $29.29 \pm 0.08$ & Herrnstein et al. (1999) \\
M87/Virgo & $31.03 \pm 0.14$ & This paper \\
\hline \hline
\end{tabular}
\end{center}
\flushleft
$^a$ Random (statistical) uncertainties only.
\end{table}

This will allow us to assign absolute distances and realistic
uncertainties to galaxies in the greater Virgo cluster environment,
while also enabling us to use the robust Virgo cluster core distance
as a stepping stone to even greater distances, all constrained to
conform with the overall distance framework. Few galaxies beyond the
largest ellipticals in the Virgo cluster have sizeable numbers of
published distance moduli available. However, many can be placed
within the 3D environment of the Virgo cluster by reference to any of
the collections of relative distance moduli compiled to date (e.g.,
Mei et al. 2007; Blakeslee et al. 2009). Most of the latter result
from relative comparisons of SBF magnitudes obtained based on
homogeneous analyses, hence constituting their own internally
consistent distance framework.

Perhaps more interesting would be an extension of the Galactic
Center--LMC--M31--(NGC 4258)--M87 reference distance ladder out to the
Fornax and Coma clusters. These clusters have been studied
extensively, but few direct, absolute distance measurements are
available. However, a not insignificant fraction of the published
distances to the Fornax and Coma clusters are relative distance moduli
with respect to that of the Virgo cluster (e.g., Blakeslee et
al. 2009; Villegas et al. 2010; and references therein). This seems an
opportune time, therefore, to extend our internally consistent
distance framework to the $\sim 20$--100 Mpc scales of Fornax and Coma
cluster-like distances. We will pursue this in Paper VII; at the
present time (2019 November 8) a NASA/ADS object search for `Fornax
cluster' returns 1869 references. A search for `Coma cluster' returned
5358 hits.

It is also interesting to briefly explore the impact of our updated,
statistically supported distance modulus to M87 on the recent
determination of that galaxy's black hole mass (Event Horizon
Telescope Collaboration et al. 2019). Most luminosity and many mass
estimates scale with $D^2$ (where $D$ is the distance), while masses
based on total densities or orbital modeling scale as $D^3$. The M87
distance modulus derived here corresponds to a distance of $16.07 \pm
1.03$ Mpc; the Event Horizon Telescope Collaboration et al. (2019)
adopted $D_{\rm M87} = 16.8 \pm 0.8$ Mpc. Although both distance
values are consistent with each other within the respective 1$\sigma$
uncertainties, our derived central value is $\sim 4$\% smaller than
theirs. Given the $D_{\rm M87}^2$ scaling of the M87 black hole mass,
this implies that with our updated distance modulus, the galaxy's
black hole mass may have to be downsized from $M_{\rm BH}^{\rm EHT} =
(6.5 \pm 0.7) \times 10^9 M_\odot$ to $M_{\rm BH}^{\rm new} = (5.9 \pm
0.6) \times 10^9 M_\odot$.

Finally, thus far we have predominantly considered the random
(statistical) uncertainties associated with our set of distance
estimates. Although we briefly discussed the nature of any differences
between the distance to M87 and that to the Virgo cluster more
broadly, the systematic uncertainties affecting our results are harder
to quantify. These uncertainties mostly originate from the calibration
choices made for any given distance indicator, many of which trace
back to the intrinsic uncertainty in the Cepheid period--luminosity
calibration. The latter has been reduced from $\sim 0.15$ mag
(Freedman et al. 2001; Macri et al. 2006) to $\sim 0.10$ mag (Freedman
\& Madore 2010), or $\sim 0.8$ Mpc at the distance of M87, in recent
years. The advent of {\sl Gaia}-based distances has the potential to
reduce these uncertainties even further (e.g., Ripepi et al. 2019; but
see Groenewegen 2018). In addition, the tie from the Cepheid distance
scale to the SBF (Tonry et al. 2001), TRGB, PNLF, or novae
calibrations used more commonly in recent years introduces an
additional systematic uncertainty, rendering the overall systematic
uncertainty affecting our results at $\sim 0.2$ mag (e.g., Bird et
al. 2010), despite all being based on a robust Local Group distance
framework.

These levels of systematic uncertainties are corroborated by studies
using multiple tracers to determine the same `best' distance to either
M87 or the Virgo cluster center. For instance, Ferrarese et al. (2000)
considered five different tracers, some covering multiple wavelengths,
resulting in a weighted average distance modulus to the center of the
Virgo cluster of $(m-M)_0 = 31.07 \pm 0.03$ mag. However, the
$1\sigma$ spread among their eight individual distance estimates is of
order 0.15 mag. Similarly, the spread among the five values published
by Tammann et al. (2000) is of order 0.11 mag.

In this context, it is instructive to consider distance estimates that
are not directly tied to any of the calibration scales we have
discussed so far. One of those stands out, in particular. The Planck
Collaboration et al. (2016) published a novel distance to the Virgo
cluster as a whole based on the Sunyaev--Zel'dovich method, $(m-M)_0 =
31.28$ mag (no uncertainty quoted). Compared with our preferred
distance modulus of 31.03 mag, this provides a good quantitative
handle on the remaining systematic uncertainties affecting secondary
distance indicators at this distance range. Note that this is a lower
limit to the error budget, since it is not yet clear whether the
difference is only caused by a difference in the zero points of the
different distance scales or if it may also be affected by the
environment. The Large Synoptic Survey Telescope will play a key role
in this context, since homogeneous multi-passband optical data will
provide an important opportunity to tie distances to Local Group and
Local Volume galaxies using the same empirical framework.

\section*{Acknowledgements}

This work was partially supported by PRIN-MIUR (2010LY5N2T), `Chemical
and dynamical evolution of the Milky Way and Local Group galaxies' (PI
F. Matteucci). This research has made extensive use of NASA's
Astrophysics Data System Abstract Service.

\end{document}